\documentclass{aastex631}

\usepackage{graphicx}
\usepackage{float}

\usepackage{array}
\usepackage{booktabs}
\usepackage{hyperref}
\usepackage{amsmath}
\begin{document}

\title{LeoNet: A Machine Learning Method for Binary Pulsar Classification}

\correspondingauthor{Wesley Armour}

\author[0009-0002-6946-7541]{Zhaocheng Gong}
\affiliation{Oxford e-Research Centre \\
Department of Engineering Science \\
University of Oxford, Oxford, UK}

\author[0000-0003-2690-6858]{Jack White}
\affiliation{Oxford e-Research Centre \\
Department of Engineering Science \\
University of Oxford, Oxford, UK}

\author[0000-0002-7865-0236]{Zeyu Yang}
\affiliation{Oxford e-Research Centre \\
Department of Engineering Science \\
University of Oxford, Oxford, UK}
\author[0000-0002-2892-8025]{Jayanta Roy}
\affiliation{National Centre for Radio Astrophysics (NCRA), Pune 411007, Maharashtra, India}
\author[0000-0003-2797-0595]{Karel Adamek}
\affiliation{Oxford e-Research Centre \\
Department of Engineering Science \\
University of Oxford, Oxford, UK}

\author[0000-0003-1756-3064]{Wesley Armour}
\affiliation{Oxford e-Research Centre \\
Department of Engineering Science \\
University of Oxford, Oxford, UK}
\email{wes.armour@oerc.ox.ac.uk}

\begin{abstract}
Binary pulsars provide valuable laboratories for testing theories of gravity, but orbital Doppler shifts complicate their detection. Fourier-domain acceleration and jerk searches address this challenge via matched filtering, but at substantial computational cost. We present LeoNet, a convolutional neural network that uses ten learnable filters to extract features of signals affected by Doppler shifts. The resulting ten-channel feature map provides a compact, lower-dimensional alternative to an explicitly sampled acceleration--jerk response grid and is analysed by a convolutional classifier to identify candidate signals. For simulated observations lasting 500\,s, LeoNet achieves a mean relative reduction in false negative rate of 55.5\% across five sampling intervals compared with the evaluated PRESTO acceleration-search configuration. TensorRT-optimised LeoNet processes each 500\,s observation in 3.44--4.37\,ms in FP32 on an NVIDIA H100 PCIe GPU across eight sampling intervals, including preprocessing, inference, and postprocessing. At a sampling interval of $128\,\mu\mathrm{s}$, its mean processing time is 3.54\,ms, compared with 1.767\,s for PRESTO FDAS on an AMD EPYC 9825 CPU with search-frequency limits of 96--1000\,Hz, corresponding to an approximately $499\times$ speedup in the measured processing time. These results suggest that LeoNet has the potential to improve detection performance, while its millisecond-scale processing time supports its use as a candidate-identification stage in real-time binary pulsar search pipelines.
\end{abstract}


\section{Introduction} \label{sec:intro}

Binary pulsar systems are of extraordinary significance, providing ideal laboratories for testing theories of gravity. Nevertheless, the discovery of new, unseen binary systems is highly challenging.

A common approach to searching for isolated pulsars is to apply a fast Fourier transform (FFT) to a dedispersed time series and examine the resulting power spectrum. The periodic pulse train concentrates signal power into narrow peaks at the pulsar's spin frequency and its harmonics, enabling candidate signals to be distinguished from noise. The inferred rotation period can then be used to fold the time series, aligning and combining successive pulses to reveal weak pulse profiles. In binary pulsar systems, however, orbital motion can cause the observed spin frequency to vary during an observation. The resulting frequency drift spreads signal power across neighbouring Fourier bins, weakening the spectral peaks and reducing sensitivity in searches that assume a constant spin frequency.

To address this issue, \cite{ransom2002fourier} proposed the Fourier Domain Acceleration Search (FDAS). This method employs multiple matched filters to convolve with the signal spectrum in the frequency domain, effectively mitigating the effects of Doppler shifts. However, FDAS is computationally expensive because it requires performing a large number of convolutions during the search process. Since the true acceleration of a pulsar is unknown prior to the acceleration search, it is necessary to explore a broad range of potential acceleration values. This results in the convolution of the input data with many trial templates, which leads to significant computational cost, which in turn limits the method's feasibility for real-time use. To improve sensitivity and enable the detection of pulsars in binary systems whose orbital motions cannot be accurately approximated by constant acceleration, the Fourier Domain "Jerk" Search (FDJS) \cite{andersen2018fourier} method has been introduced. Compared to FDAS, FDJS allows for a longer observation time, thereby increasing the likelihood of detecting fainter pulsars. However, FDJS requires templates spanning both acceleration and jerk, increasing the template count from $n$ for FDAS to $nm$, where $m$ is the number of jerk trials. If $m$ scales proportionally to $n$, the template count grows as $\mathcal{O}(n^2)$, substantially increasing the computational cost.

With the advancement of machine learning, various AI methods have been implemented to detect pulsar signals\cite{wang2019pulsar}\cite{zeng2020concat}\cite{yin2022pulsar}\cite{liu2024deep}. Solitary pulsars have distinct features that neural networks can learn, simplifying the detection process. Unfortunately, due to Doppler shifts, binary pulsars lack such obvious features, significantly increasing the difficulty of using neural networks for their detection. However, current neural network models, such as CCNN~\cite{zeng2020concat}, AR\_Net~\cite{yin2022pulsar}, and others~\cite{wang2019pulsar,liu2024deep}, are unable to correct the Doppler shift of the binary pulsar signal as effectively as FDAS, limiting their ability to detect binary pulsar signals. 

In this paper, we present LeoNet, a neural network architecture that uses a compact set of learnable Fourier-domain filters as an alternative to the larger matched-filter bank used by FDAS. This approach reduces the dimensionality of the search representation, aiming to capture relevant features of signals affected by Doppler shifts with fewer filters and lower computational cost. We compare LeoNet with FDAS in terms of search runtime and detection performance to assess its suitability for real-time binary pulsar searches.
\section{Methods} \label{sec:Data}
\subsection{Synthetic Data Generation}
In order to build a valid model for binary pulsar detection, a large amount of data is required. Currently, only a limited number of binary pulsars have been observed. As such, constructing the training set exclusively from real binary pulsar data yields a very small dataset, which is not enough for training a model effectively. Models trained on such imbalanced datasets are not well-suited for pulsar signal search.

To create a balanced training dataset, we need to generate synthetic data. We use the \colorbox{gray!20}{\texttt{fake}} SIGPROC program \cite{lorimer2011sigproc} to simulate the signal of binary pulsar systems. It is important to highlight that this will produce data with a fixed pulse profile (square pulses) in both the time and frequency domains. The amplitude of the square wave is determined by \colorbox{gray!20}{\texttt{snrpeak}}. The parameters used to generate our synthetic training dataset, along with sampling distribution used for each parameter, are shown in Table~\ref{table:synthetic}

The distributions used have been selected based on the known distributions observed in binary pulsar systems. To ensure stable training, we set a minimum snrpeak of 0.04. Including too many samples below this threshold causes the network to waste iterations fitting noise rather than learning pulsar features, which both slows convergence and reduces generalisation. For data without pulsar signals, we set the value of SNR equal to 0 to obtain pure noise samples. 

This project specifically targets millisecond pulsars. When generating our simulated dataset, we restrict spin frequencies to the range 100–1000 Hz, since every known binary pulsar spins below 1000 Hz \cite{manchester2005australia}. However, the upper bound is further constrained by the Nyquist frequency corresponding to the chosen sampling time, \(f_{\mathrm{Nyq}} = 1/(2*t_{\mathrm{samp}})\). In practice, therefore, the maximum spin frequency in the synthetic data is the minimum of 1000 Hz and \(f_{\mathrm{Nyq}}\). 

To accommodate different observing configurations, we generated synthetic datasets with observation durations of 200, 300, 400, and 500\,s and sampling intervals of 128, 256, 512, 1024, and 1310\,$\mu$s. These observing configurations were included in both the training and evaluation datasets.
\begin{table}
\centering
\begin{tabular}{c|c|c|c}
\hline
\textbf{Parameter} & \textbf{Range} & \textbf{Units} & \textbf{Distribution} \\
\hline
Spin period & [1, 10] & ms & log-uniform \\
Pulse width & [5, 50] & \% & log-uniform \\
Input SNR & [0.04, 0.1] & & log-uniform \\
Dispersion measure & [100, 1000] & cm$^{-3}$pc & log-uniform \\
Bits per sample & 8 & & \\
Number of channels & 128 & & \\
Sampling time & 128 & $\mu$s & \\
Observation time & 500 & Seconds & \\
First channel frequency & 1550 & MHz & \\
Channel offset & -0.29296875 & MHz & \\
Additional flags & -binary & & \\
Orbital period & [1, 10] & Hours & log-uniform \\
Starting orbital phase & [0, 1.0] & & uniform \\
Pulsar mass & [0.5, 5.0] & $M_\odot$ & log-uniform \\
Companion mass & [0.5, 5.0] & $M_\odot$ & log-uniform \\
Orbital eccentricity & [0, 1.0] & & uniform \\
\hline
\end{tabular}
\caption{Parameter table with range, units, and distribution (Observation time = 500s and Sampling time = 128$\mu$s).}
\label{table:synthetic}
\end{table}

\subsection{FDAS in PRESTO }

In this paper, FDAS is used as the reference method and as a control group to evaluate both the sensitivity of the neural network in detecting binary pulsar systems and the time required to complete the search. Before the acceleration search can be applied to the signals in the filterbank files, the data must be de-dispersed to obtain a time series. This is carried out using the \colorbox{gray!20}{\texttt{prepsubband}} command in PRESTO. Because the training signals are synthetic and their dispersion measures (DMs) are known a priori, only a single de-dispersion is required at the true DM, implemented with the flags \colorbox{gray!20}{\texttt{-lodm <DM value> -numdms 1}}. The resulting \colorbox{gray!20}{\texttt{.dat}} file, containing the de-dispersed time series, is then passed to the \colorbox{gray!20}{\texttt{acceleration}} routine with the option \colorbox{gray!20}{\texttt{-zmax 200}} to perform the Fourier-domain acceleration search. 

To further increase sensitivity to binary pulsars, we apply harmonic summing \cite{taylor1969two} using both \colorbox{gray!20}{\texttt{-numharm 8}} and \colorbox{gray!20}{\texttt{-numharm 16}}, which represent the most commonly adopted summing depths in pulsar searches. Finally, a detection threshold is applied to the reported significance value (\colorbox{gray!20}{\texttt{sigma}}). Candidates with $\sigma$ greater than the threshold are classified as pulsar signals, whereas those with $\sigma$ below the threshold are regarded as noise.

\subsection{Feature Extraction}
\label{sec:feature_extraction}

FDAS matches the Fourier spectrum against templates that assume constant acceleration during an observation. When the acceleration varies appreciably, the signal's phase evolution departs from this approximation, causing template mismatch and reducing the recovered matched-filter response. This motivates the use of learned features that are not restricted to a prescribed family of constant-acceleration responses.

LeoNet uses ten convolutional templates learned from the training data to represent Fourier-domain structures associated with both acceleration and jerk. Their response magnitudes form a two-dimensional feature map across frequency positions and template channels, providing a compact, lower-dimensional alternative to an explicitly sampled acceleration--jerk response grid for candidate classification. The template layer provides this representation to a subsequent classification network, which combines information across channels and neighbouring frequency positions to identify candidate signals. 

\begin{figure}
    \centering
    \includegraphics[width=\linewidth]{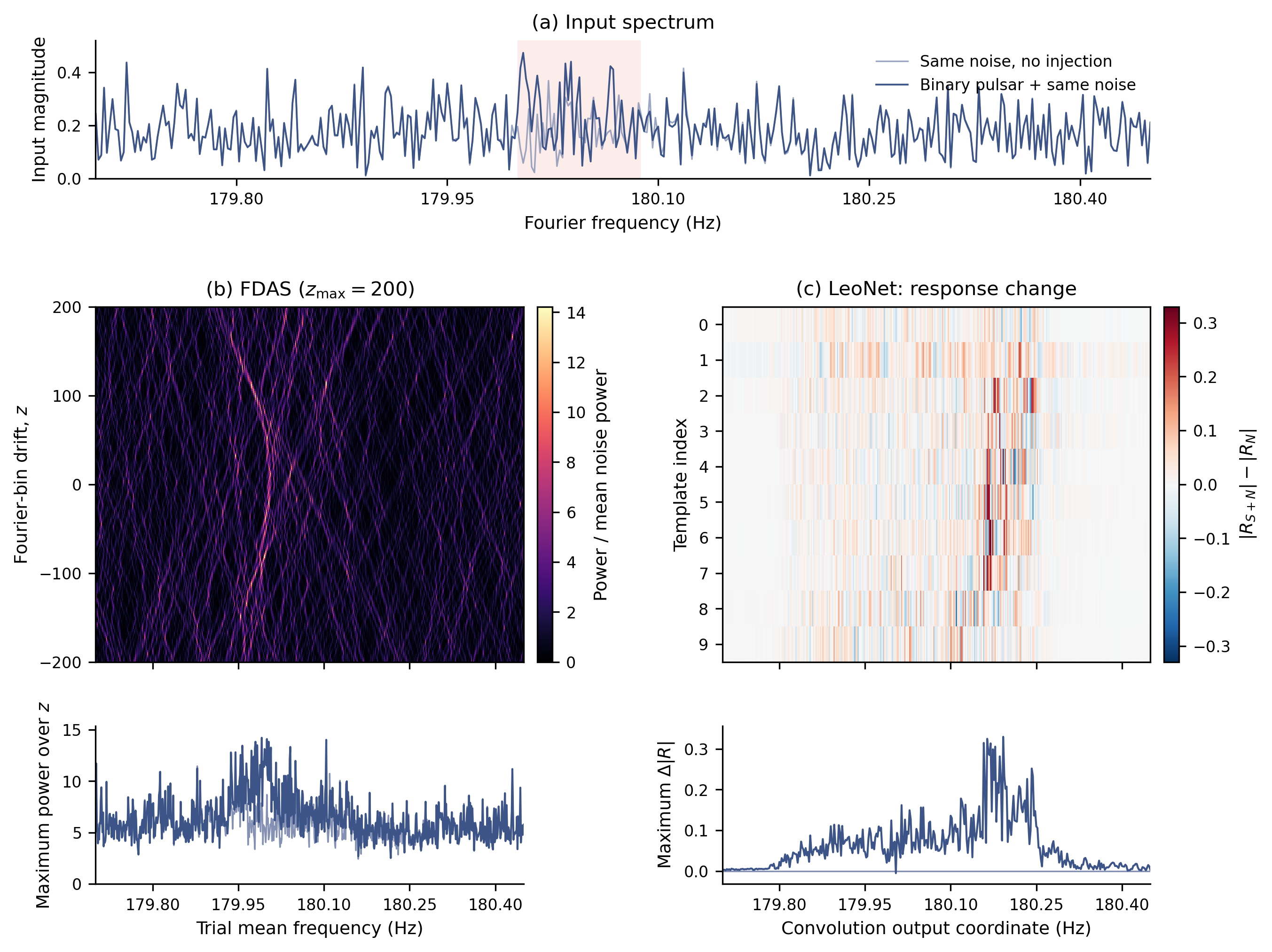}
    \caption{FDAS and LeoNet responses to a selected simulated binary pulsar injected into a Gaussian-noise filterbank. The system has a rest-frame spin period of $5.545534\,\mathrm{ms}$, orbital period of $2\,\mathrm{h}$, pulsar and companion masses of $1.4\,M_\odot$ and $1.0\,M_\odot$, eccentricity $e=0.6$, inclination $i=85^\circ$, and argument of periastron $\omega=0^\circ$. The observation has $T_{\mathrm{obs}}=500\,\mathrm{s}$ and $t_{\mathrm{samp}}=128\,\mu\mathrm{s}$, with periastron occurring 250\,s after its start. (a) Fourier amplitude spectra of the dedispersed time series before and after injection into the same noise realisation; the shaded region marks the instantaneous fundamental-frequency range. (b) Single-harmonic FDAS matched-filter power around the fundamental, normalised using the variance of the noise-only time series, for $-200\leq z\leq200$. The lower curves show the maximum power over $z$ with and without the injected signal. (c) Differences in LeoNet's ten template-response magnitudes, $\Delta|R|=|R_{S+N}|-|R_N|$, where $S$ and $N$ denote the injected signal and noise. Red and blue indicate increases and decreases, respectively. The lower curve shows the maximum difference across templates at each output position.}
    \label{fig:ffimage}
\end{figure}

Figure~\ref{fig:ffimage} illustrates this approach using a simulated binary pulsar in an eccentric orbit. The signal is generated and injected using SPECTRALib~\cite{spectralib_software} into a Gaussian-noise filterbank produced using SIGPROC. The injected and noise-only filterbanks are dedispersed using PRESTO, providing identical input time series for the two methods. The 500\,s observation spans periastron, during which the orbital acceleration and jerk vary. The FDAS response is distributed across trial frequencies and drift parameters, while the signal injection produces changes across several of LeoNet's template channels. These changes are displayed relative to the same noise-only control to make their structure easier to distinguish from the background. For this simulated observation, LeoNet detects the injected signal, whereas the PRESTO FDAS search with $z_{\max}=200$, harmonic summing up to 16 harmonics, and search-frequency limits of 96--1000\,Hz does not return a corresponding candidate. This selected example illustrates the potential of learned templates combined with a classification network to identify signals whose phase evolution is poorly represented by the constant-acceleration approximation.

The response differences in Figure~\ref{fig:ffimage} are used only for visualisation. The original template-response magnitudes form the input to the classification network, whose implementation is described in Section~\ref{sec:leonet_architecture}.

\subsection{Training Dataset Construction}

To prepare the dataset for LeoNet, we first perform de-dispersion on the raw filterbank files to obtain the time series data. The de-dispersed time series is then Fourier transformed, and the spectrum is normalised by dividing by the total number of samples. 

The number of points in the Fourier spectrum can be extremely large. For instance, with a sampling time of $128\,\mu$s and an observation length of $500\,\mathrm{s}$, the positive frequency spectrum contains $1{,}953{,}126$ points. The large input size imposes substantial computational and memory requirements, making direct processing of the full spectrum by a neural network computationally demanding. To address this, we divide the spectrum into smaller blocks. 

Choosing the block size is critical: if the blocks are too long, computation becomes inefficient, whereas blocks that are too short may not capture the full spectral features of binary pulsars, reducing detection performance. In this work, we set the block length to 4000 points as a practical compromise between computational efficiency and spectral coverage. At an observation length of 500 s, the Fourier bin width is $\Delta f = 1/T_{\mathrm{obs}} = 0.002\,\mathrm{Hz}$, so a 4000-point block spans 8 Hz in frequency. We verified on test chirp signals that this length is sufficient to preserve the characteristic structure of binary pulsars. The total number of blocks, therefore, depends on both the observation time and the sampling time. Since the maximum frequency of interest is given by $\min(1000\,\mathrm{Hz}, f_{\mathrm{Nyq}})$, where $f_{\mathrm{Nyq}} = 1/(2t_{\mathrm{samp}})$ is the Nyquist frequency, the number of blocks varies for different sampling times.

When the observation time $T_{\mathrm{obs}}$ is shorter than 500 s, the Fourier bin size increases accordingly. To ensure consistency across the dataset, we zero-pad shorter observations to 500 s before applying the Fourier transform. In addition, each block is extended with 100 overlapping points from its neighbouring regions. This overlap-padding ensures that convolution with templates, which have a length of 201 points, does not lose signal power at the block boundaries. The choice of 201-point templates corresponds to a maximum drift parameter of $z_{\mathrm{max}} = 200$, since Fourier-domain “jerk” searches have shown that most binary pulsars are detected with $|z| < 200$ (and $|w| < 600$) \cite{andersen2018fourier}. 

During training, since the fundamental spin frequency of each synthetic pulsar is known, only the block containing the corresponding frequency is extracted to construct the training and validation sets. In contrast, during inference, the pulsar frequency is unknown, so all blocks are processed by the neural network to ensure that signals can be detected regardless of their location in the spectrum.
\subsection{LeoNet Architecture}
\label{sec:leonet_architecture}

\begin{figure}
    \centering
    \includegraphics[width=1.0\linewidth]{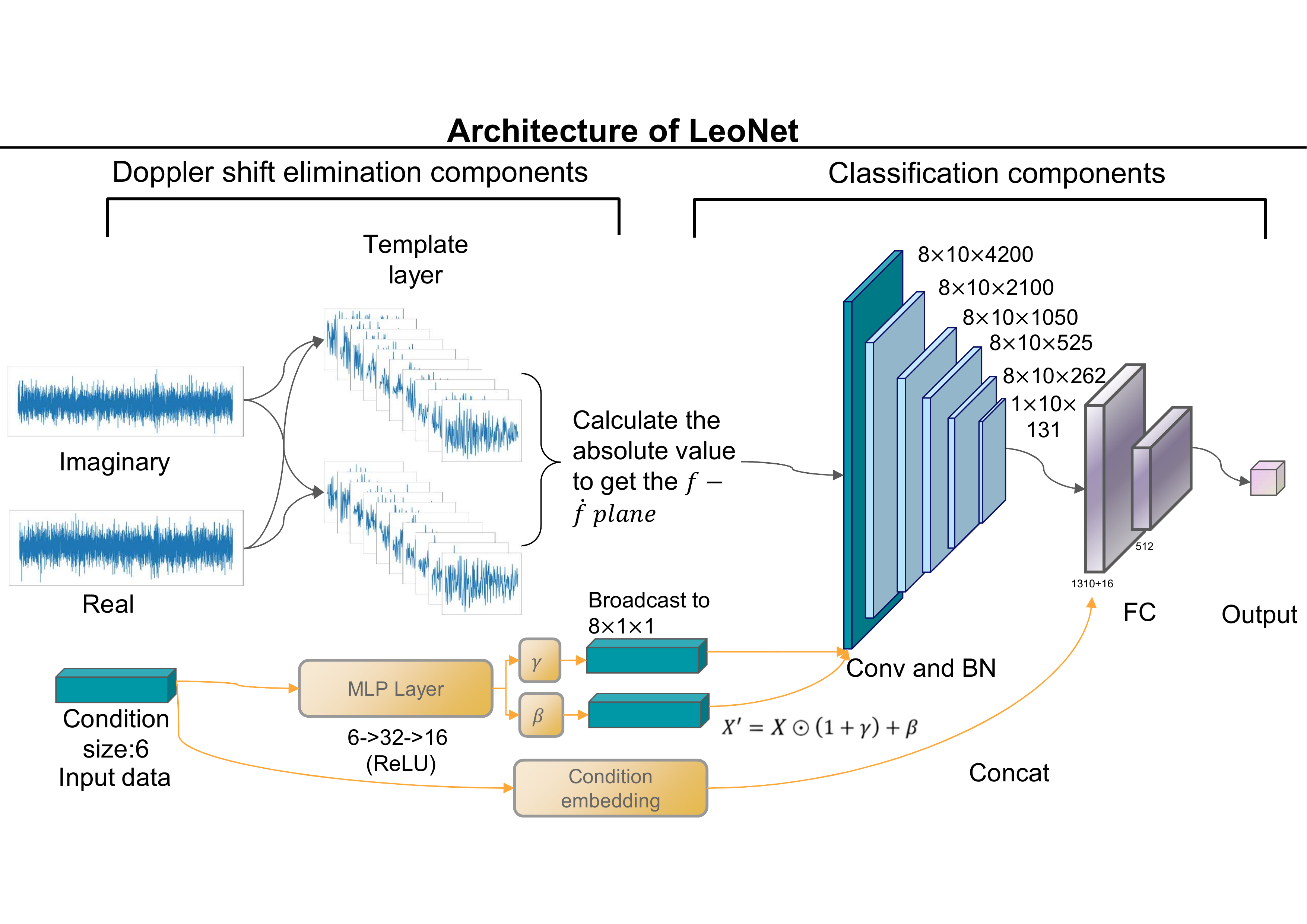}
    \caption{Architecture of LeoNet. The template layer applies ten learned convolutional templates to the complex Fourier input and combines their response magnitudes into a frequency--template feature map. The classification network processes this map using five convolutional layers and a fully connected classification head. Observing-configuration information enters through a FiLM operation after the first batch normalisation layer and through a separate conditioning embedding concatenated with the flattened CNN features.}
    \label{fig:structure of NN}
\end{figure}

LeoNet consists of two main stages: a \textit{template layer} that extracts features from the complex Fourier spectrum, and a \textit{classification network} that analyses the resulting template responses to identify candidate signals. The architecture is shown in Figure~\ref{fig:structure of NN}.

The model receives a complex-valued Fourier block containing 4200 points, represented by separate real and imaginary components, together with a six-dimensional conditioning vector describing the observing configuration. This vector contains a five-way one-hot encoding of the sampling-time category and the normalised observation duration, $\widetilde{T}_{\mathrm{obs}}=(T_{\mathrm{obs}}-100\,\mathrm{s})/(400\,\mathrm{s})$.

The template layer contains ten learnable templates, with the template count selected through a Neural Architecture Search (NAS) procedure. Each template is implemented using paired real-valued convolutional kernels of length 201. The convolutions operate along the frequency dimension with unit stride and zero-padding of 100 points on each side, preserving the input length. Their outputs are combined to form complex responses, whose magnitudes produce a $10\times4200$ frequency--template feature map. This map is supplied to the classification network as a single-channel input, with template index and frequency position forming its two spatial dimensions.

The classification network contains five two-dimensional convolutional layers: two with $5\times5$ kernels, followed by three with $3\times3$ kernels. All convolutions use unit stride and padding that preserves the spatial dimensions before pooling. The first four layers produce eight feature channels, and the final layer produces one. Each convolution is followed by batch normalisation, a ReLU activation, and max pooling with a kernel size and stride of $(1,2)$, with the FiLM operation inserted before the first ReLU as described below. Pooling progressively reduces the resolution along the frequency axis while preserving the template dimension.

A feature-wise linear modulation (FiLM) operation is applied after the first batch normalisation layer and before the ReLU activation. An MLP maps the conditioning vector through hidden layers of widths 32 and 16, followed by two linear heads that generate eight channel-wise scaling and shifting parameters, $\gamma$ and $\beta$. These parameters modulate the intermediate features according to $X'=X\odot(1+\gamma)+\beta$, with $\gamma$ and $\beta$ broadcast across the template and frequency dimensions. This provides the classification network with observing-configuration information during feature processing. 

The final CNN feature map is flattened into a 1310-dimensional vector and concatenated with a 16-dimensional embedding of the conditioning vector. This separate conditioning branch consists of a linear layer, a ReLU activation, and dropout. The resulting 1326-dimensional vector is passed through a fully connected layer with 512 hidden units, followed by ReLU activation, dropout, and a final linear layer producing a single logit. A sigmoid function converts the logit into a classification score between zero and one for the input frequency block. 

\subsection{Training Strategy}

We adopt stochastic gradient descent (SGD) with Nesterov momentum and a warm-up plus cosine annealing learning rate schedule to ensure smooth convergence. Weight decay and dropout regularisation are applied to prevent overfitting and improve generalisation.

We train LeoNet using binary cross-entropy with logits, implemented as \texttt{BCEWithLogitsLoss}. Although the training dataset is balanced, we set \texttt{pos\_weight} to 2.0, doubling the contribution of positive samples to the loss to place greater emphasis on avoiding missed pulsar signals. Detection thresholds are calibrated separately using noise samples to limit false positives during inference.

The optimiser and loss function are combined with a dynamic threshold calibration strategy. At the end of each epoch, the validation set is used to estimate both a global and configuration-specific detection threshold by computing the 99.99999th percentile of the model's scores on noise samples, where each configuration corresponds to a unique combination of $t_{\mathrm{samp}}$ and $T_{\mathrm{obs}}$. The calibrated threshold map is applied during evaluation to maintain consistent false positive control across different observing configurations while preserving sensitivity to genuine binary pulsar signals.

\begin{figure}
    \centering
    \includegraphics[width=0.9\linewidth]{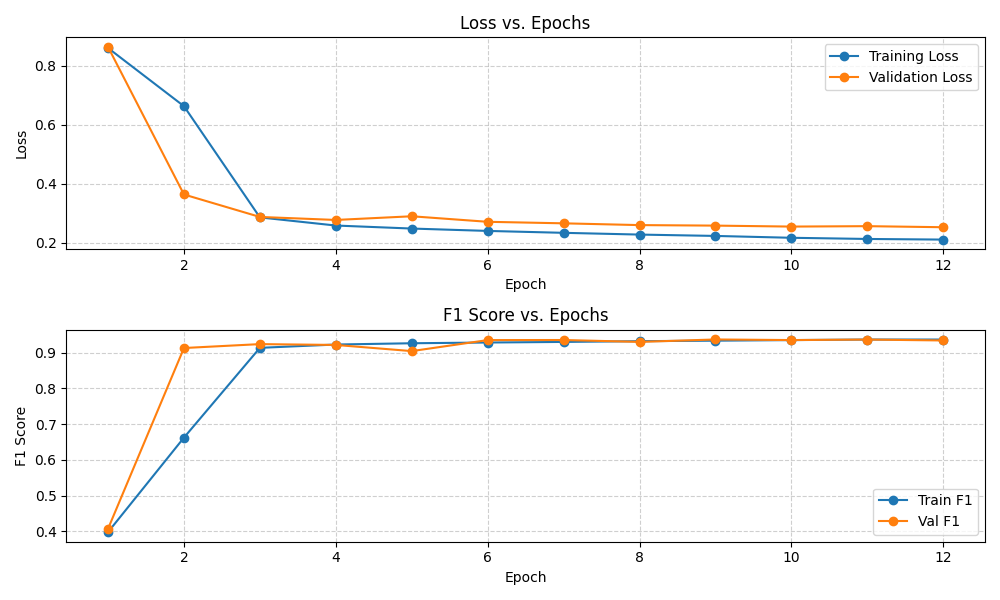}
    \caption{Training dynamics of LeoNet. The upper panel shows the training and validation loss across epochs, while the lower panel shows the evolution of the F1 score on the validation set.}
    \label{fig:training output}
\end{figure}

During training, the model performance is monitored using the F1 score, false positive rate (FPR), and false negative rate (FNR). The model weights are saved after each epoch, and the final model used for inference corresponds to the epoch with the lowest validation loss. Figure~\ref{fig:training output} shows the training dynamics, including the evolution of the loss and F1 score across epochs. The hyperparameters used in the training process are summarised in Table~\ref{tab:hyperparameters}. 
\begin{table}[htbp]
\centering
\begin{tabular}{l|l}
\hline
\textbf{Hyper-Parameter}      & \textbf{Value}       \\ \hline
Optimiser & SGD (Nesterov)       \\ 
Initial learning rate         & 0.01                 \\ 
Learning rate schedule        & Warmup(2 epochs) + CosineAnnealing  \\ 
Minimum learning rate         & 1e-4                 \\ 
Weight decay                 & 1e-3                 \\ 
Momentum                     & 0.9                  \\ 
Dropout rate                 & 0.75                 \\ 
Positive weight (pos\_weight) & 2.0            \\ 
Loss function                & BCEWithLogitsLoss    \\ 
Batch size                   & 512                  \\ 
Epochs                       & 12                   \\ \hline
\end{tabular}
\caption{Hyperparameter settings used for training LeoNet.}
\label{tab:hyperparameters}
\end{table}

\section{Evaluation}
\subsection{FDAS}
In pulsar signal searches using the FDAS, harmonic summing is applied to improve detection sensitivity for low-duty-cycle pulsars. Pulsar signals with narrow pulse profiles distribute their energy across multiple harmonics in the frequency domain. To recover these signals, FDAS coherently sums the power of several harmonics, thereby enhancing the overall SNR and improving the likelihood of detection.

In PRESTO's \texttt{accelsearch}, the number of harmonics to be summed is specified by the \texttt{-numharm} parameter. Increasing this value can improve sensitivity for pulsars with strong harmonic structures, but also increases computational cost. In practical applications, PRESTO's candidate sifting procedure uses a default significance threshold of $\sigma = 6$ to reduce false positives.

For the evaluation in this work, FDAS is executed using the \texttt{acceleration} routine in \texttt{PRESTO} with a maximum acceleration search parameter of \(-\texttt{zmax} = 200\) and harmonic summing set to \(-\texttt{numharm} = 16\). This configuration is adopted to evaluate the performance of FDAS, whilst retaining a realistic computational budget that could be incorporated into a real-time pipeline, and to provide a consistent baseline for comparison with LeoNet.
\subsection{LeoNet}
During evaluation, LeoNet analyses each observation by dividing its Fourier spectrum into multiple frequency blocks. The upper bound of the analysed frequency range depends on the sampling time and is defined as
\[
f_{\mathrm{max}} = \min(1000\,\mathrm{Hz}, f_{\mathrm{Nyq}}).
\]
Each block spans 8\,Hz in the frequency domain. Since our search targets millisecond pulsars with spin frequencies of at least 100\,Hz, we exclude the first 12 blocks below 96\,Hz, retaining the 96--104\,Hz block that contains the lower boundary of the target frequency range. Only complete blocks below $f_{\mathrm{max}}$ are evaluated, giving
\[
N_{\mathrm{blocks}}
= \left\lfloor \frac{f_{\mathrm{max}}}{8\,\mathrm{Hz}} \right\rfloor - 12.
\]
For $f_{\mathrm{max}}=1000\,\mathrm{Hz}$, each observation therefore contains 113 evaluated blocks.

LeoNet independently evaluates each block, and a binary pulsar is considered detected if any block’s output probability exceeds the defined threshold. This procedure ensures that the model can locate pulsar candidates regardless of where the harmonic appears within the spectrum.

Before inference, a threshold calibration step is performed to ensure reliable and consistent detection performance. For each combination of sampling time and observation time, over 200{,}000 pure noise blocks are evaluated using the trained network. For each observing configuration, we record the maximum pre-sigmoid logit obtained from the calibration noise samples and add a safety margin to provide an additional buffer. An excessively large margin would raise the detection threshold and could increase the number of missed signals. We therefore use a small, fixed margin of 0.1 in logit space before applying the sigmoid function to obtain the final detection threshold. The resulting threshold exceeds all outputs in the calibration noise set, but does not guarantee a zero false positive rate (FPR) on unseen data.

LeoNet searches for individual harmonic components without applying harmonic summing during evaluation. Consequently, it does not explicitly combine signal power across harmonics, which may limit its sensitivity to low-duty-cycle pulsars when individual harmonics are too weak to exceed the detection threshold.

To quantitatively assess LeoNet’s performance, several standard classification metrics are employed, including recall, precision, F1 score and FNR, defined as:
\[
\text{Recall} = \frac{\text{TP}}{\text{TP} + \text{FN}}, \qquad
\text{Precision} = \frac{\text{TP}}{\text{TP} + \text{FP}},
\]
\[
F1 = 2 \cdot \frac{\text{Precision} \cdot \text{Recall}}{\text{Precision} + \text{Recall}}, \qquad
\text{FNR} = \frac{\text{FN}}{\text{TP} + \text{FN}}
,
\]
Here, TP, TN, FP, and FN represent true positive, true negative, false positive, and false negative samples, respectively.

Precision measures the proportion of correctly identified binary pulsar signals among all detections and is crucial for large-scale pulsar surveys dominated by noise. Recall quantifies the model’s ability to identify actual pulsars, and accuracy reflects the overall correctness of the predictions. The F1 score, defined as the harmonic mean of precision and recall, provides a balanced measure of performance, especially when both false negatives and class imbalance must be carefully managed.

This evaluation framework allows LeoNet to be consistently assessed under varying sampling and observation times. For application to new telescopes or backends, the detection thresholds should be recalibrated using representative noise data, and sensitivity and false positive rates should be assessed on independent test observations.

\subsection{Evaluation on RFI-contaminated data}
\label{sec:rfi_evaluation}

To test LeoNet on RFI-contaminated data, we added simulated RFI to synthetic binary-pulsar filterbanks using SPECTRALib, applied MARS \cite{gong2026mars} for RFI mitigation, and dedispersed the resulting data at the injected DM. MARS was selected because our previous study demonstrated better RFI mitigation performance than the alternative methods evaluated in that work.

The normalisation used by MARS differs from that used for the LeoNet training data, producing offsets in the mean and Fourier-domain variance. After the FFT, we removed the mean offset by setting the DC coefficient to zero. We then matched the median blockwise variance of the real and imaginary Fourier coefficients to that of clean reference data processed with the original LeoNet preprocessing. The applied gain was

\begin{equation}
g=\sqrt{\frac{V_{\rm ref}}{V_{\rm MARS}}},
\qquad \widetilde{X}_k=gX_k,
\end{equation}

where $V_{\rm ref}$ and $V_{\rm MARS}$ are the median blockwise variances of the reference and MARS-output spectra. Since $\mathrm{Var}(gX)=g^2\mathrm{Var}(X)$, this scaling matches their spectral variance at the population-median level. For $t_{\rm samp}=128~\mu$s, $g=5.35$. The original LeoNet threshold was retained.

Before LeoNet inference, we applied a custom comb-aware birdie-zapping procedure to suppress residual periodic RFI. A comb was flagged only when at least three peaks occurred at integer-related frequencies with a peak-to-local-median ratio of $q\geq3.25$; the flagged bins were replaced with locally estimated complex Gaussian noise. The computationally dominant spectral-statistics calculations and candidate-line detection were GPU-accelerated to limit the additional preprocessing overhead. For comparison, the same MARS-mitigated time series were searched with PRESTO using the search parameters described in Section~3.1. The PRESTO \texttt{rednoise} routine was additionally applied to suppress residual low-frequency spectral noise.

Detection performance was evaluated for each simulated observation, with each observation corresponding to one input filterbank file. A TP required recovery of the Doppler-shifted target fundamental. A file was FP-contaminated if it contained at least one additional detection unrelated to the target or its allowed harmonics. This FP-file fraction measures candidate contamination and is not a negative-only false-positive rate.

\section{Discussion}
\subsection{Sensitivity Comparison of FDAS and LeoNet}
\begin{figure*}[htbp]
\centering
\gridline{
  \fig{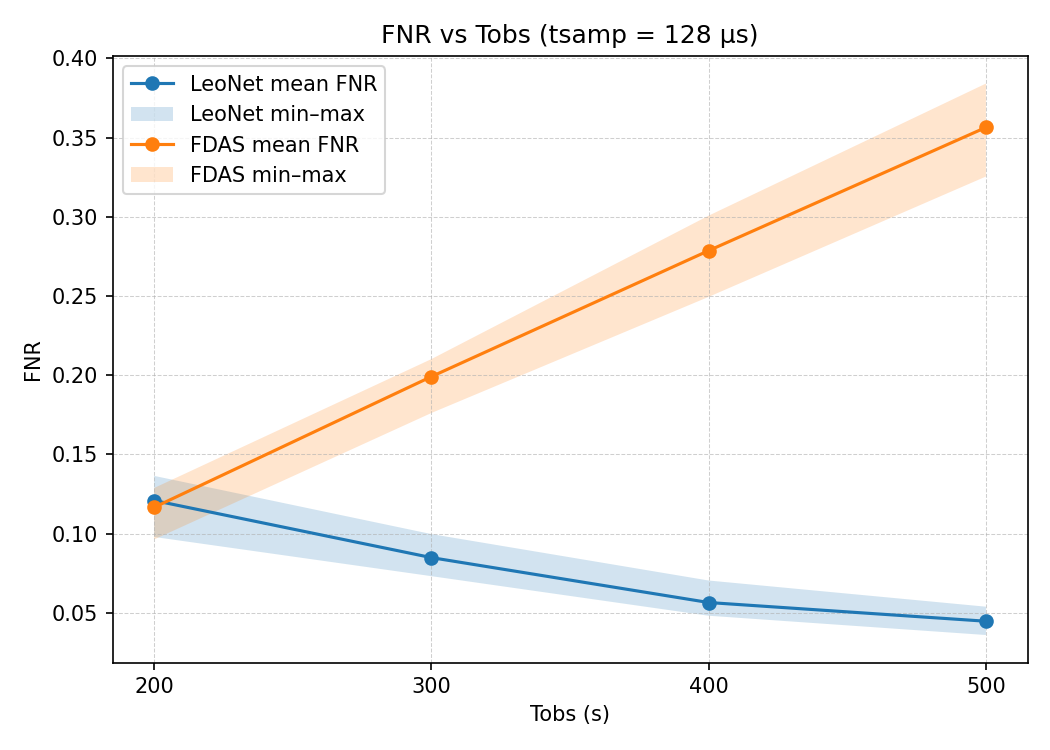}{0.32\textwidth}{$t_{\mathrm{samp}} = 128\,\mu\mathrm{s}$}
  \fig{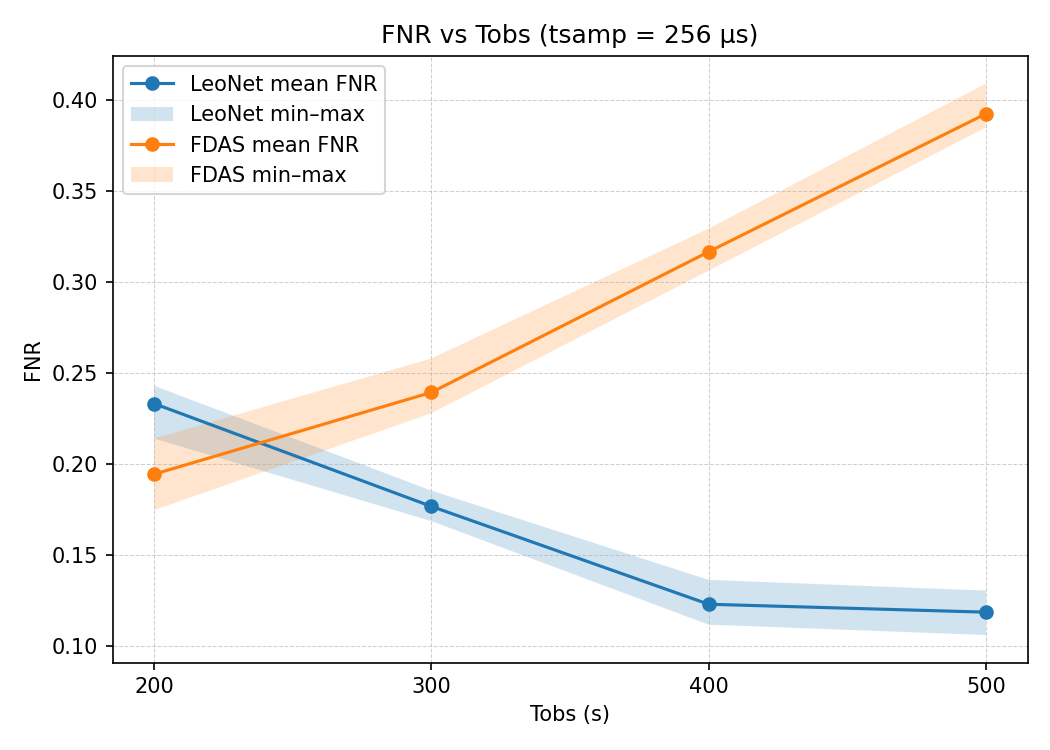}{0.32\textwidth}{$t_{\mathrm{samp}} = 256\,\mu\mathrm{s}$}
  \fig{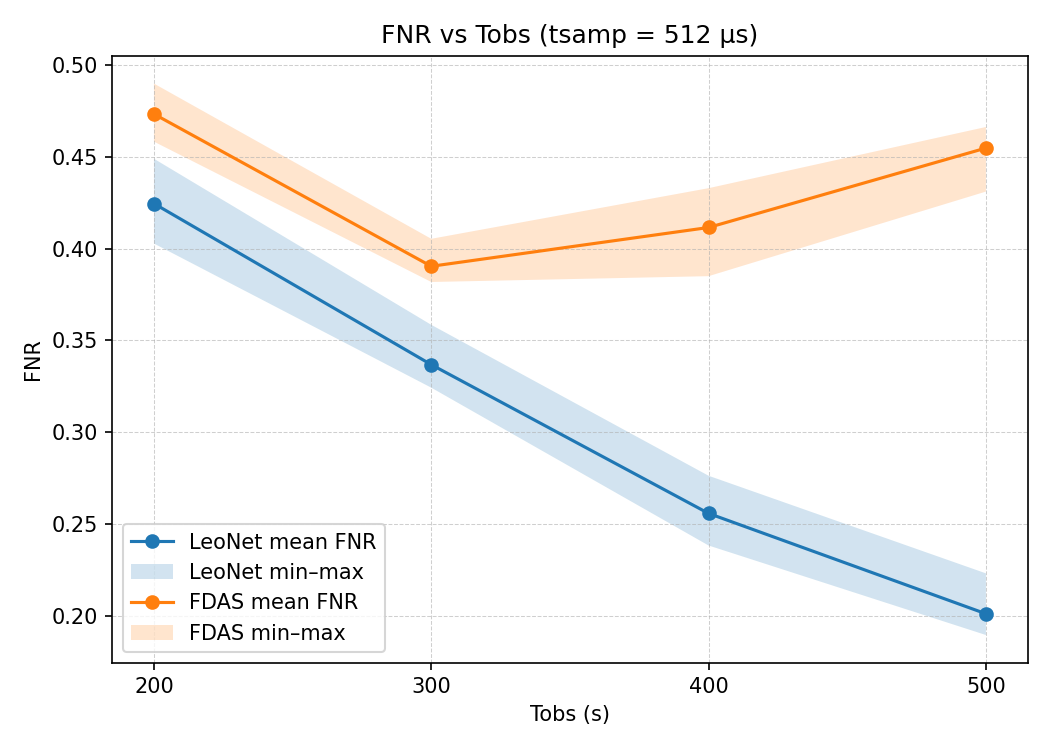}{0.32\textwidth}{$t_{\mathrm{samp}} = 512\,\mu\mathrm{s}$}
}
\gridline{
  \fig{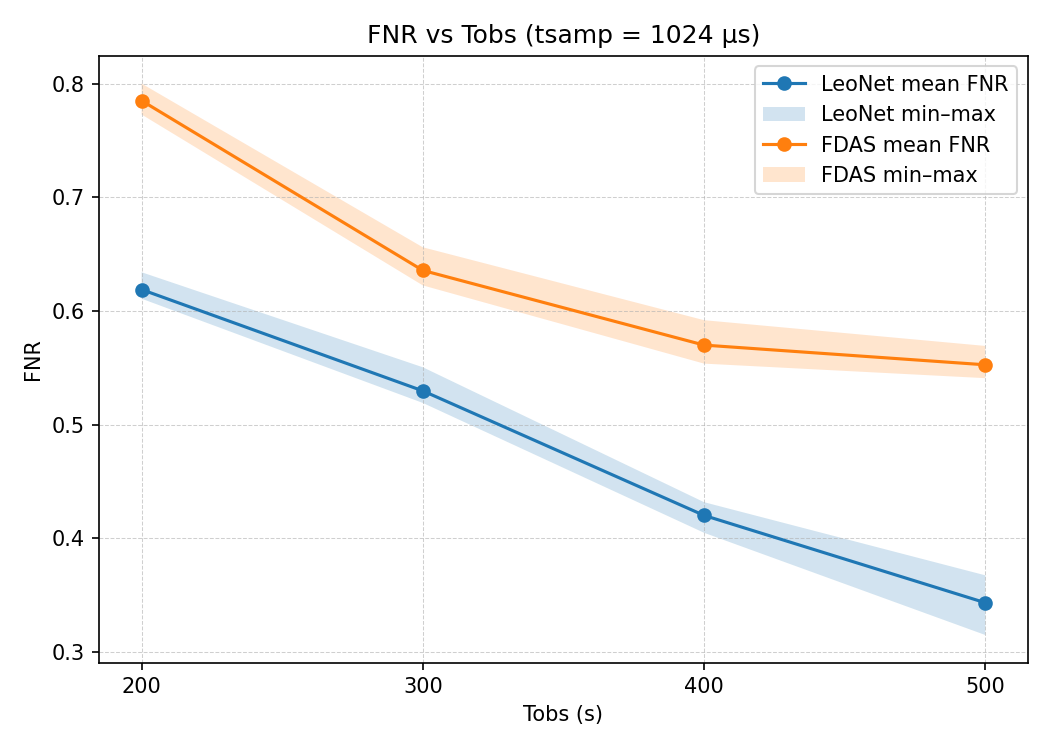}{0.32\textwidth}{$t_{\mathrm{samp}} = 1024\,\mu\mathrm{s}$}
  \fig{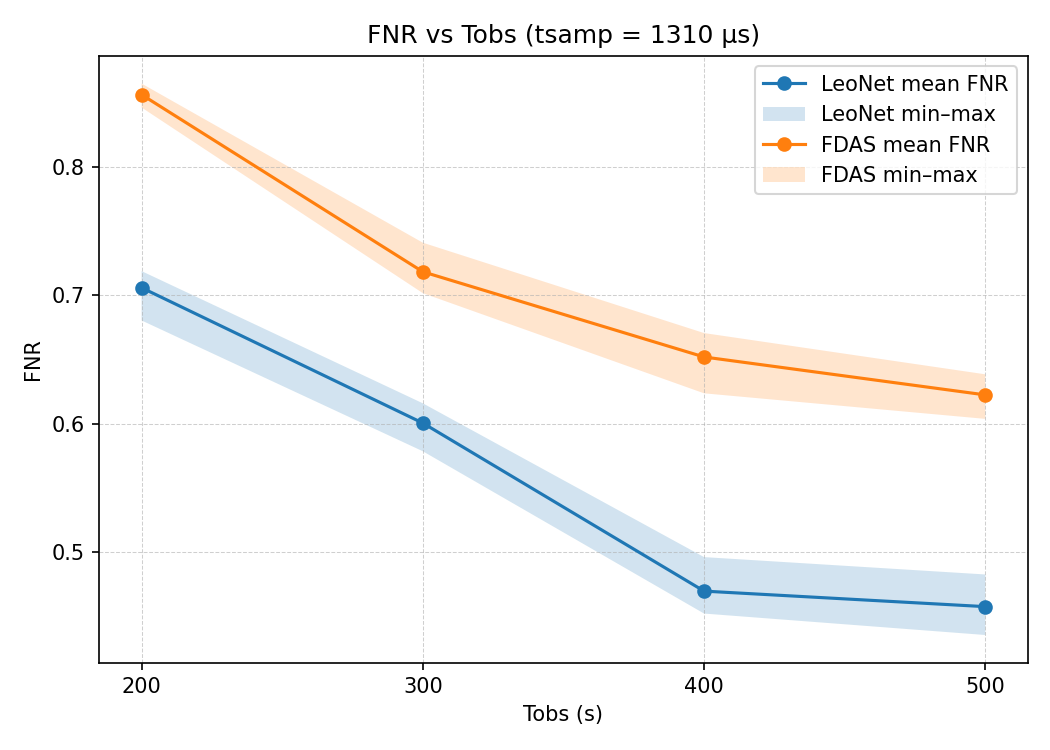}{0.32\textwidth}{$t_{\mathrm{samp}} = 1310\,\mu\mathrm{s}$}
}
\caption{FNR versus observation time for each sampling time. Curves show the mean FNR, and shaded regions represent the min--max range across multiple runs. LeoNet’s FNR generally decreases with longer integrations and remains lower than that of FDAS for $T_{\mathrm{obs}} \ge 300\,\mathrm{s}$ across all sampling times. An exception occurs at $t_{\mathrm{samp}} = 256\,\mu\mathrm{s}$ and $T_{\mathrm{obs}} = 200\,\mathrm{s}$, where FDAS achieves a slightly lower FNR due to the validity of the constant-acceleration approximation at short observation times and the sharpness of low-duty-cycle signals.}
\label{fig:fnr_tobs}
\end{figure*}
Figure~\ref{fig:fnr_tobs} compares the mean FNR of FDAS and LeoNet over observation durations of 200--500\,s for five sampling intervals. At $T_{\mathrm{obs}}=200\,\mathrm{s}$, FDAS performs comparably to LeoNet at $t_{\mathrm{samp}}=128\,\mu\mathrm{s}$ and achieves a lower mean FNR at $256\,\mu\mathrm{s}$. These results are consistent with the constant-acceleration approximation used by FDAS. For the simulated orbital periods of 1--10\,h, a 200\,s observation gives $T_{\mathrm{obs}}/P_{\mathrm{orb}}\le0.056$, satisfying the commonly used criterion $T_{\mathrm{obs}}\lesssim P_{\mathrm{orb}}/10$ \cite{andersen2018fourier}. Over such a short portion of the orbit, the line-of-sight acceleration can be approximated as constant, allowing FDAS templates to match the resulting frequency drift. 

For longer observations, changes in line-of-sight acceleration can reduce the accuracy of this approximation, while the longer integration allows more signal to accumulate. The measured FDAS trends vary with the sampling interval: its mean FNR increases with observation duration at $128$ and $256\,\mu\mathrm{s}$, decreases initially and then increases at $512\,\mu\mathrm{s}$, and decreases at $1024$ and $1310\,\mu\mathrm{s}$. LeoNet's mean FNR decreases across all five configurations, although the additional improvement between 400 and 500\,s is small at $256$ and $1310\,\mu\mathrm{s}$. At every tested duration from 300 to 500\,s, LeoNet achieves a lower mean FNR than FDAS across all five sampling intervals. 

Tables~\ref{tab:leonet_500s} and~\ref{tab:fdas_500s} quantify the comparison at $T_{\mathrm{obs}}=500\,\mathrm{s}$. LeoNet achieves lower FNR and higher recall and F1 scores at every sampling interval under the adopted detection thresholds. Averaging the relative changes calculated separately for the five sampling intervals gives a 55.5\% reduction in FNR and a 26.2\% increase in F1 score. The relative FNR reduction ranges from 26.5\% at $1310\,\mu\mathrm{s}$ to 87.4\% at $128\,\mu\mathrm{s}$. Both methods show higher FNRs at coarser sampling intervals. We next examine the properties of signals detected by FDAS and those detected only by LeoNet. 

\begin{table*}[htbp] 
\centering 
\begin{minipage}[t]{0.48\linewidth} 
\centering 
\caption{Evaluation metrics of LeoNet at $T_{\mathrm{obs}}=500\,\mathrm{s}$ using the calibrated detection thresholds.} 
\label{tab:leonet_500s} 
\small 
\setlength{\tabcolsep}{4pt} 
\begin{tabular}{l|cccc} 
\hline 
$t_{\mathrm{samp}}$ ($\mu$s) & FNR & Recall & F1 & Precision \\ 
\hline 
128  & 0.0448 & 0.9552 & 0.9771 & 1.000 \\ 
256  & 0.1184 & 0.8816 & 0.9371 & 1.000 \\ 
512  & 0.2010 & 0.7990 & 0.8882 & 1.000 \\ 
1024 & 0.3434 & 0.6566 & 0.7925 & 1.000 \\ 
1310 & 0.4576 & 0.5424 & 0.7031 & 1.000 \\ 
\hline 
\end{tabular} 
\end{minipage} 
\hfill 

\begin{minipage}[t]{0.48\linewidth} 
\centering 
\caption{Evaluation metrics of FDAS using PRESTO's \texttt{accelsearch} with \texttt{-numharm 16} and $\sigma=6$ at $T_{\mathrm{obs}}=500\,\mathrm{s}$.} 
\label{tab:fdas_500s} 
\small 
\setlength{\tabcolsep}{4pt} 
\begin{tabular}{l|cccc} 
\hline 
$t_{\mathrm{samp}}$ ($\mu$s) & FNR & Recall & F1 & Precision \\ 
\hline 
128  & 0.3564 & 0.6436 & 0.7829 & 1.000 \\ 
256  & 0.3922 & 0.6078 & 0.7560 & 1.000 \\ 
512  & 0.4548 & 0.5452 & 0.7055 & 1.000 \\ 
1024 & 0.5527 & 0.4473 & 0.6181 & 1.000 \\ 
1310 & 0.6224 & 0.3776 & 0.5481 & 1.000 \\ 
\hline 
\end{tabular} 
\end{minipage} 
\end{table*} 

\begin{figure*}[htbp] 
    \centering 
    \includegraphics[width=\textwidth]{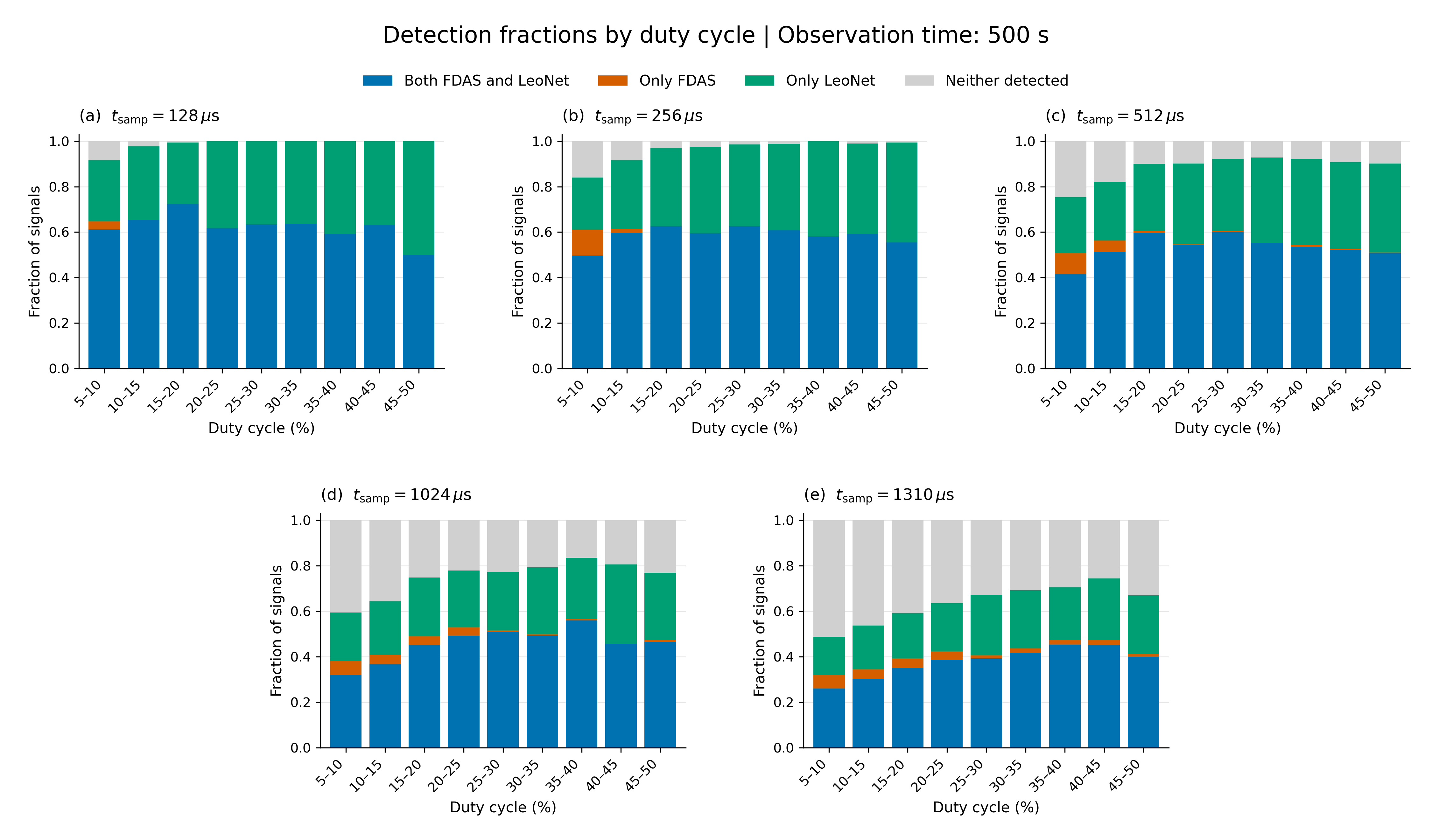} 
    \caption{Detection outcomes as a function of duty cycle at $T_{\mathrm{obs}}=500\,\mathrm{s}$ for sampling intervals of (a) 128, (b) 256, (c) 512, (d) 1024, and (e) 1310\,$\mu$s. Within each duty-cycle bin, the stacked bars show the fractions detected by both FDAS and LeoNet (blue), only by FDAS (orange), only by LeoNet (green), and by neither method (grey). The total detection fractions for LeoNet and FDAS are the sums of the blue and green components and of the blue and orange components, respectively. FDAS uses PRESTO's \texttt{accelsearch} with \texttt{-numharm 16} and $\sigma=6$.} 
    \label{fig:duty_cycle} 
\end{figure*}

Figure~\ref{fig:duty_cycle} separates the detection outcomes by duty cycle at $T_{\mathrm{obs}}=500\,\mathrm{s}$ for all five sampling intervals. Each bar distinguishes signals detected by both methods, detected only by FDAS, detected only by LeoNet, and missed by both. The fraction detected only by LeoNet exceeds the fraction detected only by FDAS in every displayed duty-cycle bin. Thus, LeoNet achieves a higher detection fraction within each bin, although some signals remain detectable only by FDAS. 

At $t_{\mathrm{samp}}=128\,\mu\mathrm{s}$, LeoNet recovers nearly all signals with duty cycles above approximately 15\%, while its missed detections are concentrated towards lower duty cycles. At $256$ and $512\,\mu\mathrm{s}$, the FDAS-only detections are also most apparent in the lowest duty-cycle bins. At $1024$ and $1310\,\mu\mathrm{s}$, a larger fraction of signals is missed by both methods, including at higher duty cycles. LeoNet continues to recover more signals in each bin, but small FDAS-only fractions persist over a broader duty-cycle range. 

\begin{figure*}[htbp] 
    \centering 
    \includegraphics[width=\textwidth]{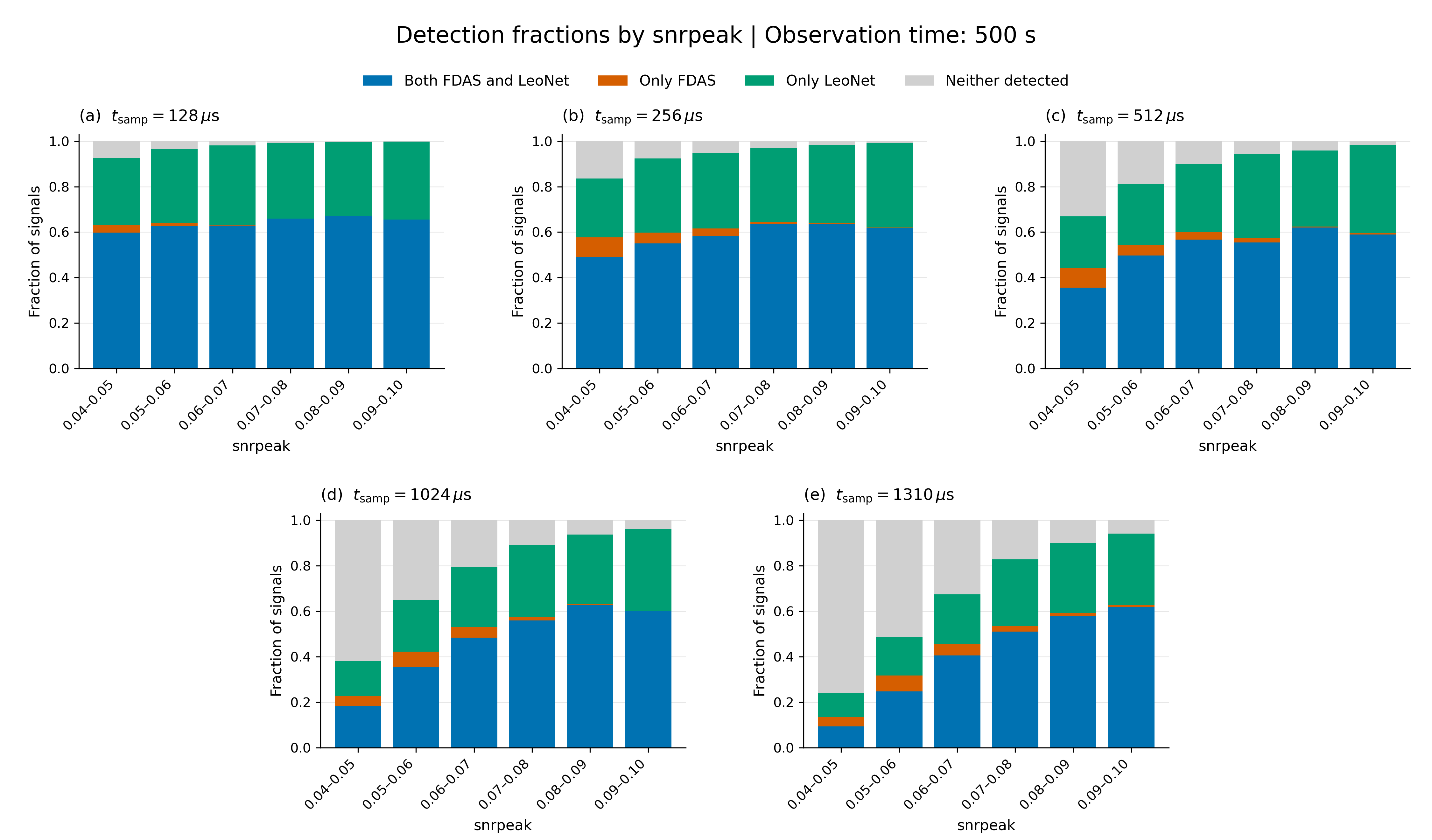} 
    \caption{Detection outcomes as a function of the simulation input parameter \texttt{snrpeak} at $T_{\mathrm{obs}}=500\,\mathrm{s}$ for sampling intervals of (a) 128, (b) 256, (c) 512, (d) 1024, and (e) 1310\,$\mu$s. Within each bin, the stacked bars show the fractions detected by both FDAS and LeoNet (blue), only by FDAS (orange), only by LeoNet (green), and by neither method (grey). LeoNet's detection fraction is the sum of the blue and green components; the FDAS detection fraction is the sum of the blue and orange components. FDAS uses PRESTO's \texttt{accelsearch} with \texttt{-numharm 16} and $\sigma=6$.} 
    \label{fig:snr_comparison} 
\end{figure*}

To examine the dependence on input signal strength, Figure~\ref{fig:snr_comparison} presents the detection outcomes as a function of \texttt{snrpeak} at $T_{\mathrm{obs}}=500\,\mathrm{s}$. Here, \texttt{snrpeak} is the simulation input parameter controlling the amplitude of the injected pulse profile. LeoNet's detection fraction generally increases with this parameter. At $t_{\mathrm{samp}}=128\,\mu\mathrm{s}$, LeoNet already recovers a large fraction of signals in the lowest \texttt{snrpeak} bin, with recovery approaching unity at higher values. At coarser sampling intervals, particularly $1024$ and $1310\,\mu\mathrm{s}$, a much larger fraction of the weakest injected signals is missed by both methods. Recovery improves as \texttt{snrpeak} increases, although some signals remain undetected even in the highest bin. 

The fraction detected only by LeoNet exceeds the fraction detected only by FDAS in every displayed \texttt{snrpeak} bin. LeoNet therefore achieves a higher detection fraction throughout the input signal-strength range examined here, including the weakest-signal bins where both methods have reduced recovery. Nevertheless, FDAS-only detections remain, predominantly towards the lower end of the \texttt{snrpeak} range. 

Harmonic summing may contribute to the FDAS-only detections at low duty cycles and low input signal strengths. Narrow pulse profiles distribute signal power across multiple harmonics, which PRESTO's \texttt{accelsearch} can combine through incoherent harmonic summing to improve detection significance \cite{andersen2018fourier}. LeoNet evaluates individual frequency blocks without combining power across harmonics, which may limit its recovery of signals whose individual harmonics are weak. 

At coarser sampling intervals, a small number of high-duty-cycle signals are also detected only by FDAS. These cases indicate a residual limitation of LeoNet under the adopted detection thresholds, despite its higher overall detection fraction. 
\subsection{Pulscan vs LeoNet}

We compared LeoNet with the GPU implementation of \textit{Pulscan} \cite{white2025pulscan}, an unmatched-boxcar search tool developed for real-time binary pulsar searches. The comparison used synthetic observations with $T_{\mathrm{obs}}=500\,\mathrm{s}$ and $t_{\mathrm{samp}}=128\,\mu\mathrm{s}$. We first applied a periodicity search and excluded observations with significant detections. Of the original 7,200 signal-containing observations, we removed 2,978 whose first-listed candidate had a reported significance of $\sigma\geq 6$. Both methods were evaluated on the same remaining 4,222 signal-containing observations. We tested both methods on 5,400 samples of pure noise to evaluate their performance on such data and check for false positives.

We evaluated \textit{Pulscan} at detection-significance thresholds of $\sigma=6$, 7, and 8. Table~\ref{tab:pulscan_thresholds} summarises the resulting counts and rates. At $\sigma=6$, \textit{Pulscan} recovered 4,073 of the 4,222 target signals, corresponding to an FNR of 3.5291\%, but also produced 1,719 false positives. Such false detections increase the workload for subsequent candidate screening and verification in a real-time pipeline. Increasing the threshold to $\sigma=7$ reduced the false-positive count to seven, while recovering 3,929 target signals and yielding an FNR of 6.9398\%. At $\sigma=8$, no false positives were observed, but the number of recovered signals decreased to 3,845, corresponding to an FNR of 8.9294\%. These results quantify the trade-off between signal recovery and false detections over the tested thresholds.

\begin{table}[htbp]
    \centering
    \caption{
    Detection results for the GPU implementation of \textit{Pulscan} at $T_{\mathrm{obs}}=500\,\mathrm{s}$ and $t_{\mathrm{samp}}=128\,\mu\mathrm{s}$. TP and FN refer to the 4,222 target signals. The final column reports the FPR.
    }
    \label{tab:pulscan_thresholds}
    \begin{tabular}{c|rrr|rr}
        \hline
        $\sigma$ & TP & FN & FNR (\%) & FP & FPR (\%) \\
        \hline
        6 & 4073 & 149 & 3.5291 & 1719 & 31.8333 \\
        7 & 3929 & 293 & 6.9398 &    7 &  0.1296 \\
        8 & 3845 & 377 & 8.9294 &    0 &  0.0000 \\
        \hline
    \end{tabular}
\end{table}

We adopted $\sigma=7$ for the comparison with LeoNet, as it substantially reduced false positives relative to $\sigma=6$ while retaining more target signals than $\sigma=8$.

Using its calibrated detection threshold, LeoNet recovered 4,033 of the same 4,222 target signals and missed 189, yielding an FNR of 4.48\%. No false positives were observed among the same 5,400 noise-only samples, giving an observed FPR of 0\%. Compared with \textit{Pulscan} at $\sigma=7$, LeoNet recovered 104 more signals overall, reducing the FNR by 35.5\% while producing fewer false positives.

\begin{figure}[htbp]
    \centering
    \includegraphics[width=0.6\linewidth]{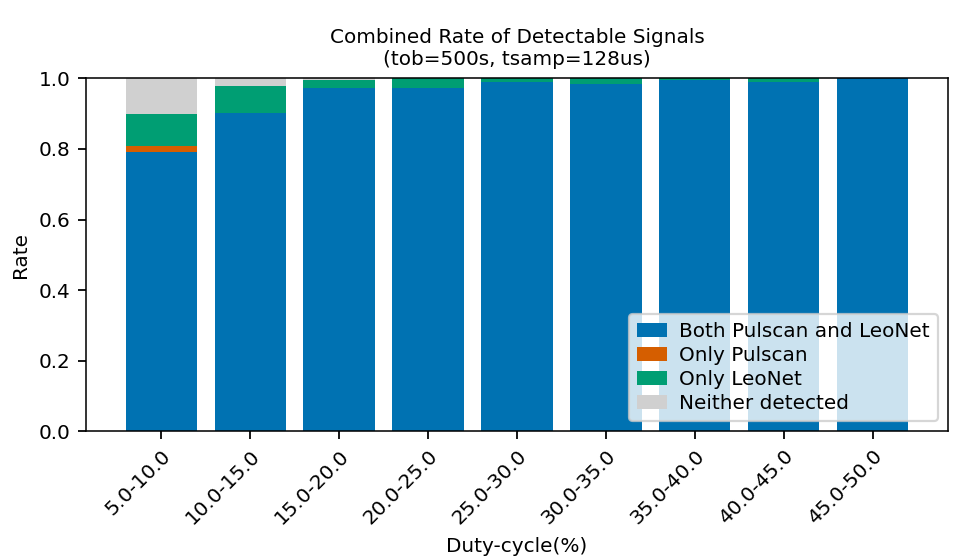}
    \caption{
    Detection outcomes as a function of duty cycle for
    $T_{\mathrm{obs}}=500\,\mathrm{s}$ and
    $t_{\mathrm{samp}}=128\,\mu\mathrm{s}$.
    Each stacked bar shows the fractions of signals in the
    corresponding duty-cycle bin detected by both
    \textit{Pulscan} and LeoNet (blue), only by
    \textit{Pulscan} (orange), only by LeoNet (green),
    or by neither method (grey).
    }
    \label{fig:pulscan_duty_cycle}
\end{figure}

Figure~\ref{fig:pulscan_duty_cycle} shows the detection outcomes across duty-cycle bins, distinguishing signals detected by both methods, only by LeoNet, only by \textit{Pulscan}, or by neither. In the $5\%<D\le10\%$ duty-cycle bin, 27 signals were detected only by \textit{Pulscan}, compared with 130 detected only by LeoNet. No signals were detected exclusively by \textit{Pulscan} in the remaining duty-cycle bins. These \textit{Pulscan}-only detections in the 5--10\% duty-cycle bin are likely due to the sensitivity gain provided by the four-harmonic summing applied by default in \texttt{pulscan\_gpu}.

\subsection{Robustness to RFI-contaminated data}
\label{sec:rfi_discussion}

We evaluated LeoNet and PRESTO on MARS-mitigated observations at five sampling intervals using the procedure described in Section~\ref{sec:rfi_evaluation}. Each dataset initially
contained 1,000 observations, and the first $T_{\rm obs}=500\,\mathrm{s}$ of each observation were searched.

At $t_{\rm samp}=128\,\mu\mathrm{s}$, four observations with severe RFI contamination were excluded because their time series were unusable for pulsar searches after RFI mitigation. At $t_{\rm samp}=256$ and $512\,\mu\mathrm{s}$, respectively, 17 and 19 processed outputs were excluded because they contained non-finite or extreme values, or fell outside the expected output distribution. All 1,000 observations were retained at both $t_{\rm samp}=1024\,\mu\mathrm{s}$ and $t_{\rm samp}=1310\,\mu\mathrm{s}$. Both methods were evaluated on the same retained observations at each sampling interval.

Table~\ref{tab:mars_rfi_all_ts} compares LeoNet and PRESTO on the retained MARS-mitigated observations across all five sampling intervals.

\begin{table*}[htbp]
    \centering
    \small
    \setlength{\tabcolsep}{4pt}
    \caption{
    File-level results for LeoNet and PRESTO after MARS RFI mitigation at $T_{\rm obs}=500\,\mathrm{s}$. $N$ is the number of retained observations. Recall and FP-file percentages are calculated using $N$.
    An FP-contaminated file contains at least one detection unrelated to the injected target or its allowed harmonics. The FP-file fraction measures candidate contamination, rather than a negative-only false-positive rate.
    }
    \label{tab:mars_rfi_all_ts}
    \begin{tabular}{r r r r r r r r r}
        \hline
        & & & \multicolumn{3}{c}{MARS--LeoNet}
        & \multicolumn{3}{c}{MARS--PRESTO} \\
        \cline{4-6}\cline{7-9}
        $t_{\rm samp}$ & $g$ & $N$
        & TP & Recall & FP files
        & TP & Recall & FP files \\
        ($\mu$s) & & &
        & (\%) & $n$ (\%)
        & & (\%) & $n$ (\%) \\
        \hline
        128  & 5.35   & 996
        & 911 & 91.47 & 44 (4.42)
        & 762 & 76.51 & 89 (8.94) \\
        256  & 5.3500 & 983
        & 814 & 82.81 & 40 (4.07)
        & 701 & 71.31 & 24 (2.44) \\
        512  & 5.3430 & 981
        & 650 & 66.26 & 71 (7.24)
        & 540 & 55.05 & 43 (4.38) \\
        1024 & 5.4192 & 1000
        & 536 & 53.60 & 51 (5.10)
        & 416 & 41.60 & 36 (3.60) \\
        1310 & 5.5229 & 1000
        & 437 & 43.70 & 66 (6.60)
        & 329 & 32.90 & 29 (2.90) \\
        \hline
    \end{tabular}
\end{table*}

At $t_{\rm samp}=128\,\mu\mathrm{s}$, LeoNet recovered 911 target signals after variance matching, compared with 762 for PRESTO, yielding recalls of 91.47\% and 76.51\%, respectively. LeoNet therefore achieved 14.96 percentage points higher recall. It also produced fewer FP-contaminated files: 44 (4.42\%) compared with 89 (8.94\%) for PRESTO,
a reduction of 4.52 percentage points.

LeoNet achieved higher recall than PRESTO at the remaining sampling intervals, with improvements of 11.50, 11.21, 12.00, and 10.80 percentage points at $t_{\rm samp}=256$, 512, 1024, and $1310\,\mu\mathrm{s}$, respectively. However, LeoNet also produced a higher
FP-file fraction in these configurations. Thus, although LeoNet recovered more target signals at every tested sampling interval, its lower candidate contamination relative to PRESTO was limited to the $128\,\mu\mathrm{s}$ configuration.

The frequency coverage varied with sampling interval. Because LeoNet processes complete 8-Hz blocks, its input blocks covered 96--1000\,Hz at $t_{\rm samp}=128$ and $256\,\mu\mathrm{s}$, 96--976\,Hz at $512\,\mu\mathrm{s}$, 96--488\,Hz at $1024\,\mu\mathrm{s}$, and 96--376\,Hz at $1310\,\mu\mathrm{s}$. For false-positive statistics, candidates were counted only when their reported frequencies lay within the target evaluation range, from $100\,\mathrm{Hz}$ to $f_{\mathrm{max}}=\min(1000\,\mathrm{Hz},f_{\mathrm{Nyq}})$, where $f_{\mathrm{Nyq}}$ is the Nyquist frequency.

Some residual LeoNet false positives may arise from harmonics of RFI below this evaluation range. For example, an RFI signal with a fundamental frequency of $60\,\mathrm{Hz}$ may produce harmonics at 120 and 180\,Hz. Although the fundamental is outside the evaluation range, these harmonics are inside it and may be identified by LeoNet as pulsar candidates. Our comb-aware birdie zapping requires at least three detectable peaks belonging to the same harmonic series. It may therefore leave such interference unflagged when only one or two peaks satisfy the detection criterion. At coarser sampling intervals, the lower Nyquist frequency can further reduce the number of observable harmonics available to identify a comb.

PRESTO may report the same interference at a different frequency. Its harmonic summing combines power from multiple harmonics and may report a candidate at the underlying fundamental, such as $60\,\mathrm{Hz}$ in this example. That candidate falls below our evaluation range and is therefore excluded from the false-positive count. This can explain why an RFI signal produces a counted false positive for LeoNet but not for PRESTO. The FP-file fractions in the table nevertheless reflect all counted false-positive candidates across the evaluated observations, so these individual cases do not imply that PRESTO always produces fewer FP-contaminated files.

\subsection{Timing}
The inference efficiency of LeoNet can be improved using specialised optimisations. We evaluate three execution backends: vanilla PyTorch, TorchScript followed by \texttt{torch.jit.optimize\_for\_inference} (JIT-OFI), and Torch-TensorRT using its TorchScript frontend. JIT-OFI applies inference-oriented graph optimisations, while Torch-TensorRT compiles supported operations into GPU-optimised TensorRT engines.

We assess performance using the measured processing time per observation. The benchmarks use 500\,s dedispersed time series with sampling intervals of 49.152, 64, 81.92, 128, 256, 512, 1024, and 1310\,$\mu$s. For each sampling interval, the same input observation is used for LeoNet and PRESTO. LeoNet is evaluated on three NVIDIA GPUs: the RTX 3090, A100 PCIe 40\,GB, and H100 PCIe. During inference, the batch size is set to 128.

The LeoNet timings include all processing stages from the dedispersed time series to the detection records: FFT computation and normalisation, frequency-block extraction, FiLM conditioning, batch assembly, network inference, sigmoid evaluation, thresholding, and detection-record construction. The input time series is resident in GPU memory before timing begins. Transfers between CPU and GPU memory are excluded in both directions. This measurement targets integration into a GPU-based processing pipeline, in which upstream operations such as dedispersion have already produced data which would be resident in GPU memory. Disk reads, model loading and compilation, and result-file writing are excluded.

For comparison, the Fourier Domain Acceleration Search is performed using PRESTO's \texttt{accelsearch} on an AMD EPYC 9825 processor, with \texttt{-zmax 200} and harmonic summing up to 16 harmonics. To select the number of CPU threads, we performed a thread sweep over 1, 2, 4, 8, 16, 32, 64, 96, 144, 192, and 288 threads (\texttt{-ncpus}). Using 16 threads gave the lowest mean processing time for all eight sampling intervals, so all PRESTO timings reported below use 16 threads. To restrict the nominal search band to that considered by LeoNet, PRESTO is run with \texttt{-flo 96 -fhi 1000}. These parameters specify the search-frequency limits for the highest harmonic in PRESTO's harmonic-summing convention. For sampling intervals whose Nyquist frequency is below 1000\,Hz, the available frequency coverage is correspondingly reduced. The input and output files used by PRESTO are placed in a memory-backed filesystem (\texttt{tmpfs}) to reduce storage-related overhead.

For both methods, each configuration is warmed up with 10 executions, followed by 50 timed executions of the same observation. We report the arithmetic mean of these 50 measurements, excluding the warm-up executions.

\begin{figure}
    \centering
    \includegraphics[width=1.0\linewidth]{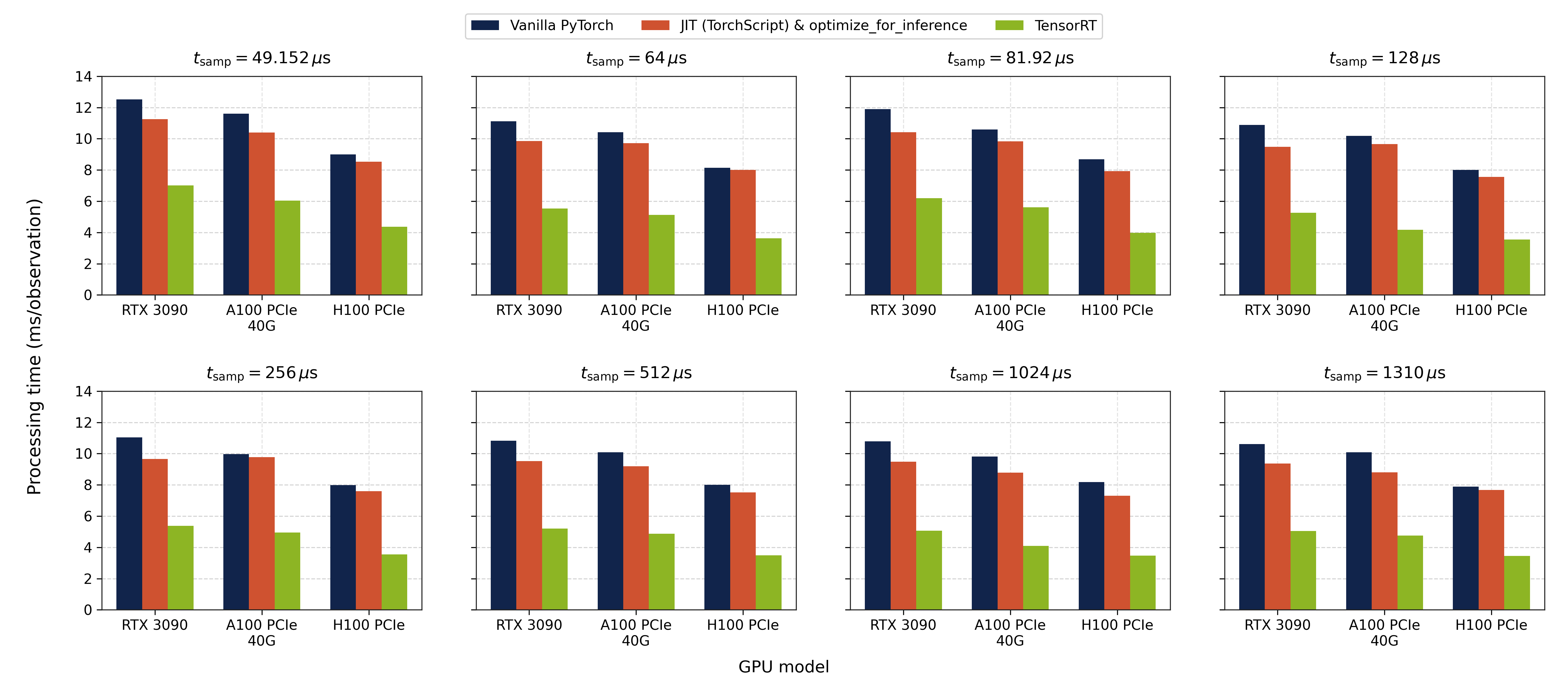}
    \caption{Mean processing time per 500\,s observation for LeoNet using vanilla PyTorch, JIT-OFI, and TensorRT on three NVIDIA GPUs. Each panel corresponds to a different sampling interval. All configurations use FP32 with the backends' default TF32 settings and a batch size of 128 frequency blocks. Bars show arithmetic means over 50 timed executions following 10 warm-up executions. Timings include FFT and normalisation, block extraction and FiLM conditioning, inference, and postprocessing. CPU--GPU transfers, disk I/O, and model loading and compilation are excluded.}
    \label{fig:perf_comparison}
\end{figure}

\begin{figure}[H]
    \centering
    \includegraphics[width=0.9\linewidth]{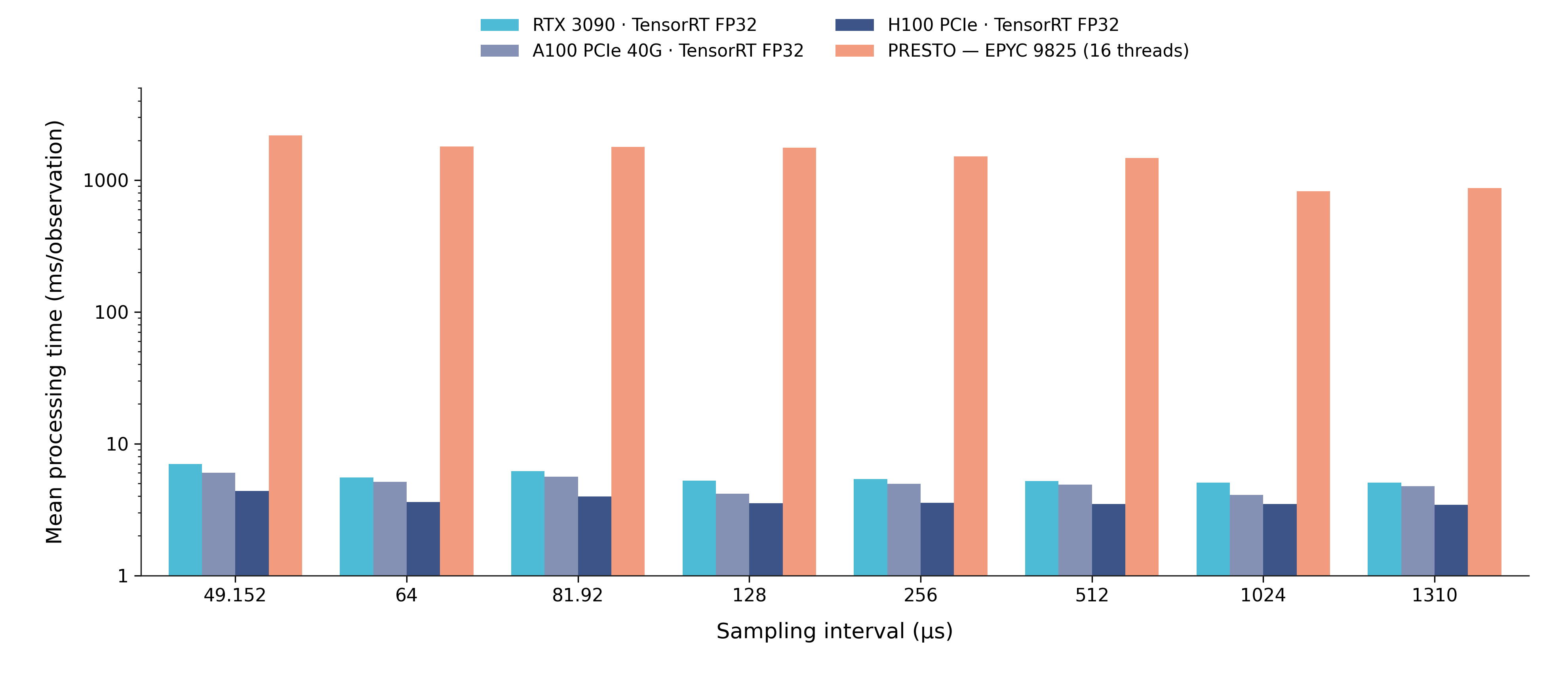}
    \caption{Mean processing time per 500\,s observation for LeoNet with TensorRT in FP32 and PRESTO on an AMD EPYC 9825 processor. The vertical axis is logarithmic. PRESTO uses \texttt{-zmax 200}, \texttt{-numharm 16}, and \texttt{-flo 96 -fhi 1000}; the latter specify the highest-harmonic search-frequency limits. Frequency coverage is limited by the input Nyquist frequency. LeoNet timings include preprocessing, inference, and postprocessing, but exclude CPU--GPU transfers and disk I/O. PRESTO timings use a warmed memory-backed filesystem (\texttt{tmpfs}) and retain process-startup and file-operation costs. Each bar represents the arithmetic mean of 50 timed executions after 10 warm-up executions.}
    \label{fig:presto_comparison}
\end{figure}

Figure~\ref{fig:perf_comparison} compares the three LeoNet backends. Across the 24 GPU--sampling-interval combinations, JIT-OFI and
TensorRT achieve geometric-mean speedups of $1.09\times$ and $2.12\times$, respectively, relative to vanilla PyTorch. Figure~\ref{fig:presto_comparison} compares TensorRT-accelerated LeoNet with PRESTO. On the H100 PCIe, LeoNet requires 3.44--4.37\,ms per observation across the tested sampling intervals, whereas PRESTO requires 822.74--2184.18\,ms on the AMD EPYC 9825. At a sampling interval of 128\,$\mu$s, the measured mean times are 3.54\,ms and 1767.34\,ms, respectively. As illustrated in the above figures, the network approach offers a substantial improvement in processing speed compared to FDAS. This highlights its considerable potential for real-time binary pulsar searches and demonstrates its effectiveness in handling high-resolution astronomical data efficiently.

\section{Conclusion}
We have presented LeoNet, a neural network architecture for the efficient detection of binary pulsar signals. Its template layer uses a compact set of learnable Fourier-domain filters to extract features of signals affected by Doppler shifts, providing an alternative to the larger matched-filter bank used by FDAS and FDJS.

On synthetic data without RFI at $T_{\rm obs}=500\,\mathrm{s}$, LeoNet achieved a mean relative reduction in FNR of approximately 55.5\% compared with FDAS across the five evaluated sampling intervals. At $t_{\rm samp}=128\,\mu\mathrm{s}$, LeoNet also reduced the FNR by 35.5\% relative to the GPU implementation of \textit{Pulscan} operating at $\sigma=7$. 

Evaluation on synthetic RFI-contaminated observations further demonstrated the applicability of LeoNet after MARS mitigation and the additional preprocessing described in this work. LeoNet achieved higher recall than PRESTO across all five sampling intervals, with improvements of 10.80--14.96 percentage points. Its FP-file fraction was lower than that of PRESTO at $t_{\rm samp}=128\,\mu\mathrm{s}$, but higher at the remaining sampling intervals. Residual RFI therefore remains a source of candidate contamination despite the improvement in target recovery.

Inference optimisation further reduced the processing time of LeoNet. Across the three NVIDIA GPU types and eight sampling intervals tested, TensorRT achieved a geometric-mean speedup of $2.12\times$ relative to vanilla PyTorch in the FP32 configuration. For a 500\,s observation sampled at $128\,\mu\mathrm{s}$, LeoNet required a mean processing time of 3.54\,ms on the NVIDIA H100 PCIe, compared with 1.767\,s for PRESTO FDAS using 16 threads on an AMD EPYC 9825 CPU with search-frequency limits configured to 96--1000\,Hz. Under these timing conditions, the results demonstrate the low per-observation processing cost of LeoNet and support its potential for integration into GPU-based real-time binary pulsar search pipelines.

LeoNet remains limited in recovering some short-duty-cycle signals and rejecting residual periodic RFI. Future work will investigate incorporating harmonic information, improving residual RFI rejection, extending the training data to cover a wider range of pulsar parameters, and validating the approach on real observations from different telescopes and backends.

\vspace{5mm}
\software{SIGPROC, SIGPYPROC, MARS, PRESTO, SPECTRALIB}
\clearpage
\newpage
\bibliography{reference}{}
\bibliographystyle{aasjournal}



\end{document}